\documentclass{aa}  

\usepackage{graphicx}
\usepackage{etoolbox, siunitx}

\usepackage{gensymb}
\usepackage{orcidlink}

\setcitestyle{notesep={ }}

\usepackage{txfonts}
\begin{document}

\title{TOI-1259Ab: A Warm Jupiter Orbiting a K-dwarf White-Dwarf Binary is on a Well-aligned Orbit}

   %\subtitle{Sub title}

\author{
Hugo Veldhuis\orcidlink{0009-0009-6875-4128}\inst{1}
\and
Juan I. Espinoza-Retamal\orcidlink{0000-0001-9480-8526}\inst{2,3,1}
\and
Gudmundur Stefansson\orcidlink{0000-0001-7409-5688}\inst{1}
\and
Alexander P. Stephan\orcidlink{0000-0001-8220-0548}\inst{4}
\and
David V. Martin\orcidlink{0000-0002-7595-6360}\inst{5}
\and
David Bruijne\orcidlink{0009-0006-0218-6269}\inst{1}
\and
Suvrath Mahadevan\orcidlink{0000-0001-9596-7983}\inst{6,7}
\and
Joshua N. Winn\orcidlink{0000-0002-4265-047X}\inst{8}
\and
Cullen H. Blake\orcidlink{0000-0002-6096-1749}\inst{9}
\and
Fei Dai\orcidlink{0000-0002-8958-0683}\inst{10}
\and
Rachel B. Fernandes\orcidlink{0000-0002-3853-7327}\inst{6,7}\thanks{President's Postdoctoral Fellow}
\and
Evan Fitzmaurice\orcidlink{0000-0003-0199-9699}\inst{6,7,11}
\and
Eric B. Ford\orcidlink{0000-0001-6545-639X}\inst{6,7,11,12}
\and
Mark R. Giovinazzi\orcidlink{0000-0002-0078-5288}\inst{13}
\and
Arvind F.\ Gupta\orcidlink{0000-0002-5463-9980}\inst{14}
\and
Samuel Halverson\orcidlink{0000-0003-1312-9391}\inst{15}
\and
Te Han\orcidlink{0000-0002-7127-7643}\inst{16}
\and
Daniel Krolikowski\orcidlink{0000-0001-9626-0613}\inst{17}
\and
Joe Ninan\orcidlink{0000-0001-8720-5612}\inst{18}
\and
Cristobal Petrovich\orcidlink{0000-0003-0412-9314}\inst{19}
\and
Paul Robertson\orcidlink{0000-0003-0149-9678}\inst{16}
\and
Arpita Roy\orcidlink{0000-0001-8127-5775}\inst{20}
\and
Christian Schwab\orcidlink{0000-0002-4046-987X}\inst{21}
\and
Ryan Terrien\orcidlink{0000-0002-4788-8858}\inst{22}
}

% -------------------------------------------
% -------------------------------------------
% -------------------------------------------
\institute{Anton Pannekoek Institute for Astronomy, University of Amsterdam, Science Park 904, 1098 XH Amsterdam, The Netherlands
\and
Instituto de Astrof\'isica, Pontificia Universidad Cat\'olica de Chile, Av. Vicu\~na Mackenna 4860, 782-0436 Macul, Santiago, Chile
\and
Millennium Institute for Astrophysics, Santiago, Chile
\and
Department of Physics and Astronomy, Vanderbilt University, TN 37235, USA
\and
Department of Physics and Astronomy, Tufts University, 574 Boston Avenue, Medford, MA 02155
\and
Department of Astronomy \& Astrophysics, The Pennsylvania State University, 525 Davey Laboratory, University Park, PA 16802, USA
\and
Center for Exoplanets and Habitable Worlds, The Pennsylvania State University, 525 Davey Laboratory, University Park, PA 16802, USA
\and
Department of Astrophysical Sciences, Princeton University, 4 Ivy Lane, Princeton, NJ 08544, USA
\and
Department of Physics and Astronomy, University of Pennsylvania, 209 South 33rd Street, Philadelphia, PA 19104, USA
\and
Institute for Astronomy, University of Hawai’i, 2680 Woodlawn Drive, Honolulu, HI 96822, USA
\and
Institute for Computational and Data Sciences, The Pennsylvania State University, University Park, PA 16802, USA
\and
Center for Astrostatistics, 525 Davey Laboratory, The Pennsylvania State University, University Park, PA 16802, USA
\and
Department of Physics and Astronomy, Amherst College, Amherst, MA 01002, USA
\and
U.S. National Science Foundation National Optical-Infrared Astronomy Research Laboratory, 950 N.\ Cherry Ave., Tucson, AZ 85719, USA
\and
Jet Propulsion Laboratory, California Institute of Technology, 4800 Oak Grove Drive, Pasadena, CA 91109, USA
\and
Department of Physics \& Astronomy, The University of California, Irvine, Irvine, CA 92697, USA
\and
Steward Observatory, University of Arizona, 933 N. Cherry Ave, Tucson, AZ 85721, USA
\and
Department of Astronomy and Astrophysics, Tata Institute of Fundamental Research, Homi Bhabha Road, Colaba, Mumbai 400005, India
\and
Department of Astronomy, Indiana University, Bloomington, IN 47405, USA
\and
Astrophysics \& Space Institute, Schmidt Sciences, New York, NY 10011, USA
\and
School of Mathematical and Physical Sciences, Macquarie University, Balaclava Road, North Ryde, NSW 2109, Australia
\and
Carleton College, One North College Street, Northfield, MN 55057, USA
}

\date{}

\abstract{The evolution of one member of a stellar binary into a white dwarf has been proposed as a mechanism that triggers the formation of close-in gas giant planets. The star's asymmetric mass loss during the AGB stage gives it a ``kick'' that can initiate Eccentric Lidov-Kozai oscillations, potentially causing a planet around the secondary star to migrate inwards and perturbing the eccentricity and inclination of its orbit. Here we present a measurement of the stellar obliquity of TOI-1259Ab, a gas giant in a close-in orbit around a K star with a white dwarf companion about  1650\,au away. By using the NEID spectrograph to detect the Rossiter-McLaughlin effect during the planetary transit, we find the sky-projected obliquity to be $\lambda = 6^{+21}_{-22}\,\degree$. When combined with estimates of the stellar rotation period, radius, and projected rotation velocity, we find the true 3D obliquity to be $\psi = 24^{+14}_{-12}\,\degree$ ($\psi < 48\degree$ at 95\% confidence), revealing that the orbit of TOI-1259Ab is well aligned with the star's equatorial plane. Because the planet's orbit is too wide for tidal realignment to be expected, TOI-1259Ab might have formed quiescently in this well-aligned configuration. Alternatively, as we show with dynamical simulations, Eccentric Lidov-Kozai oscillations triggered by the evolution of the binary companion are expected to lead to a low obliquity with a probability of about $\sim14\%$.}

\maketitle

% ------------------------------------------
% ------------------------------------------
% ------------------------------------------
\section{Introduction}
Since the discovery of the first exoplanet around a solar-type star \citep{Mayor95}, the existence of gas giants in short-period orbits has been puzzling. The theories that have been proposed to explain these close-in planets typically fall into three categories \citep[see e.g.,][]{HJ_review_dawson2018}. The first category is high-eccentricity migration, in which a planet that formed beyond the ice line of the protoplanetary disk is launched onto a highly eccentric orbit through planet-planet scattering \citep[e.g.,][]{Rasio96,beauge2012}, Eccentric Lidov-Kozai (ELK) oscillations \citep[e.g.,][]{kozai_cycles, lidov_cycles, Naoz2016}, or other secular interactions \citep[e.g.,][]{Wu11,Petrovich15}, after which tidal friction circularizes and shrinks the orbit. The second category is disk-driven migration, in which the planet spirals inward as a consequence of gravitational tidal interactions with the gaseous protoplanetary disk \citep[e.g.,][]{Goldreich80,Lin86}. The third category is in situ formation, in which the planet forms close to the star, despite previous theoretical arguments that this should not occur \citep[e.g.,][]{Batygin16,Boley16}.

Insights into the formation and evolution of planetary systems can be gained from studying the stellar obliquity ($\psi$)---the angle between a star's spin axis and the planet's orbital axis \citep[see e.g.,][]{albrecht_dawson_winn_2022}. For the planets in the Solar System, these angles are low, with a $\sim 7^\circ$ inclination of the solar rotation axis as seen from Earth \citep{Sun_obliquity_Beck_2005}. In contrast, observations of the Rossiter-McLaughlin (RM) effect \citep{mclaughlin1924, rossiter1924} have revealed a wide range of obliquities seen in exoplanetary systems including well-aligned, polar, and even retrograde orbits. Recently, \cite{Albrecht2021} found hints of a preponderance of nearly polar orbits in the obliquity distribution of exoplanet systems, although its significance is still in question \citep[e.g.,][]{Siegel2023,Dong23,juan_polar_neptunes}.

Recent work by \citet{Stephan+2024} suggested that the evolution to the white dwarf (WD) phase of a stellar companion can increase the probability of high-eccentricity migration of gas giant planets. Through a dynamical "kick" from asymmetric mass loss during the star's AGB phase \citep[e.g.,][]{fregeau_WD_kick, el_Badry_WD_kick}, the orbit of the companion could become tilted with respect to the orbital plane of the planet, a condition that allows for ELK oscillations \citep[e.g.,][]{Wu03,Fabrycky07,Naoz+2012}. This formation scenario could be tested with a population of obliquity measurements of systems with a WD companion. This population, however, is currently quite meager, as only three obliquity measurements of such systems have been reported. Furthermore, at the time of the obliquity measurement, two of their WD companions had not been detected, so their novel architectures have not been discussed.
%the first two did not discuss their novel architecture as their companions were not yet discovered.

Here, we present a measurement of a low obliquity for TOI-1259Ab, a close-in gas giant orbiting a bright K star with a WD companion \citep{discovery_Martin, Fitzmaurice_2022_WD}. We detected the RM effect using precise radial velocity (RV) observations with the NEID spectrograph on the WIYN\footnote{The WIYN Observatory is a joint facility of the NSF's National Optical-Infrared Astronomy Research Laboratory, Indiana University, the University of Wisconsin-Madison, Pennsylvania State University, Purdue University and Princeton University.} 3.5 m telescope at Kitt Peak in Arizona. Having a relatively large orbital separation \citep[$a/R_{\star} \approx 12$;][]{discovery_Martin}, the planet's orbit is too wide for tidal realignment to be expected, making it seem likely that the system formed with a low obliquity. To further investigate the WD-kick formation scenario proposed by \cite{Stephan+2024}, we performed ELK simulations of the system and compared the measured obliquity value of TOI-1259Ab with the distribution of obliquities predicted by the simulations.

This paper is structured as follows. In Section \ref{sec:The TOI-1259 System}, we present the TOI-1259 system, and we discuss the observations in Section \ref{sec:observations}. In Section \ref{sec:RM_fit_and_analysis}, we discuss our analysis of the RM effect of TOI-1259Ab and accompanying datasets. In Section \ref{sec:Discussion}, we put our results in context with previous obliquity measurements, and we show that a low obliquity is a possible but somewhat improbable outcome of the WD kick formation scenario. We conclude with a summary of our findings in Section \ref{sec:Conclusion}.

% -----------------------------------------------------
% -----------------------------------------------------
% -----------------------------------------------------
\section{The TOI-1259 System}
\label{sec:The TOI-1259 System}
TOI-1259 is a binary system, consisting of a $0.744 ^{+0.064}_{-0.059}\, \textrm{M}_\odot$ K-dwarf primary star ($T_{\rm{eff}}=4775\,\pm\,100$ K) and a $0.561\,\pm\,0.021\,\textrm{M}_{\odot}$ WD companion \citep{discovery_Martin}. Using data from the Transiting Exoplanet Survey Satellite \citep[TESS;][]{Ricker2015} combined with RV follow-up observations, \cite{discovery_Martin} discovered a transiting close-in gas giant around the primary star, with an orbital period of $P = 3.4779780^{+0.0000019}_{-0.0000017}$ days, a mass of $M_p =0.441^{+0.049}_{-0.047}\, \textrm{M}_{\textrm{J}}$, and a radius of $R_p = 1.022^{+0.030}_{-0.027}\,\textrm{R}_{\textrm{J}}$. Recent work by \citet{saidel2025atmospheric} found evidence for an extended escaping atmosphere of the planet, but concluded that atmospheric mass loss has not significantly altered the properties of the planet.

\citet{discovery_Martin} derived the age of the TOI-1259 system using two independent methods. First, the TESS lightcurves reveal photometric modulation of the primary star with a period of $\sim28$ days, presumably the stellar rotation period. Using gyrochronology the age of TOI-1259A was calculated to be $4.8^{+0.7}_{-0.8}$ Gyr. Second, since WDs steadily cool over time, a measurement of the effective temperature is a proxy for the WD's age. With this method, \citet{discovery_Martin} determined TOI-1259B's age, finding that, broadly speaking, the ages for TOI-1259A and TOI-1259B are consistent, as one would expect for a co-eval binary.

Follow-up spectroscopy of the WD by \citet{Fitzmaurice_2022_WD} revealed it to be a DA-type WD, characterized by strong hydrogen Balmer absorption lines as its only prominent spectral features \citep[see, e.g.,][]{Saumon2022}. This classification, along with cooling models, further constrained its age to $4.0^{+1.0}_{-0.4}$ Gyr. TOI-1259B also shows no signs of atmospheric pollution by heavy elements. Such pollution is observed in approximately 25–50\% of WDs \citep[e.g.,][]{WD_pol_Zuckerman_2010, WD_pol_zuckerman2014, WD_pol_koester2014} and is thought to trace the accretion of planetary debris \citep[e.g.,][]{Aguilera2025}. However, the absence of pollution does not preclude past or present planets around TOI-1259B. For example, WD 1856+534 hosts a transiting giant planet yet shows no detectable metal lines in its spectrum \citep{Vanderburg2020}.

% -----------------------------------------------------
% -----------------------------------------------------
\section{Observations}\label{sec:observations}

% -----------------------------------------------------
\subsection{Spectroscopic Data}
We observed one transit of TOI-1259Ab using the NEID spectrograph \citep{NEID_design_schwab2016} on the WIYN 3.5m telescope at Kitt Peak Observatory in Arizona. NEID is an environmentally stabilized \citep{stefansson2016,Robertson_environment}, high-resolution ($R\approx110,000$) spectrograph covering a broad wavelength range, from 3800 to 9300 $\AA$. We observed the transit on the night of April 22nd 2022, between 05:50 and 10:26 UTC. During the observation, the target rose from airmass 1.99 to 1.51. We obtained 26 spectra of the host star during the transit, each with an exposure time of 600 seconds. The RVs were extracted using the NEID Data Reduction Pipeline (DPR; version 1.40)  \footnote{\url{https://neid.ipac.caltech.edu/docs/NEID-DRP/}}. The average signal-to-noise ratio of the final spectra in NEID order index 102 (865 nm) is 8.7 per 1D extracted pixel, leading to an average RV precision of $\sim8$ m s$^{-1}$. We see no correlation between the derived RVs and the H$\alpha$, Na D I, or Na D II stellar activity indices provided by the DRP. The extracted RVs are shown in Figure \ref{fig:RM_fits} and can be found in Appendix \ref{app:rvs}.

In order to better constrain the parameters of the planet and its orbit, we used 19 archival RV observations taken with the SOPHIE spectrograph \citep{sophie_specs}. SOPHIE is a fiber-fed high-resolution ($R\approx75,000$) échelle spectrograph mounted on the 1.93 m telescope of Observatoire de Haute-Provence in France. These earlier observations were used to confirm the planetary nature of the transiting object \citep{discovery_Martin} and were performed between June and July 2020.

% ----------------------------------------------------------
\subsection{Photometric Data}
\label{sec:TESS_text}
We analyzed the spectroscopic data in combination with
photometric observations. In particular, we included the 2-minute TESS light curves from a total of 30 sectors---namely Sectors 14, $17-21$, $24-26$, 40, 41, 47, 48, $50-55$, $58-60$, $73-75$, and $77-81$---produced with the TESS Science Processing Operations Center (SPOC) pipeline \citep{spoc}. This collection includes 21 more sectors than were analyzed by \citet{discovery_Martin}. We searched for, downloaded, and combined all the light curves using the \texttt{lightkurve} package \citep{Lightkurve}. Additionally, we detrended the light curve using a Matern-3/2 Gaussian Process (GP) with the \texttt{juliet} \citep{juliet} code, which uses \texttt{celerite} \citep{celerite} for the GP and \texttt{batman} \citep{batman} to model the transits. From the \texttt{juliet} analysis, we obtained the detrended light curve and the ephemerides of the planet, both of which were used in the final joint fit discussed in Section \ref{sec:RM_fit_and_analysis}. 

% ----------------------------------------------------------
% ----------------------------------------------------------
% ----------------------------------------------------------
\section{Joint Analysis}
\label{sec:RM_fit_and_analysis}

\begin{figure*}
\centering
\includegraphics[width=0.92\linewidth]{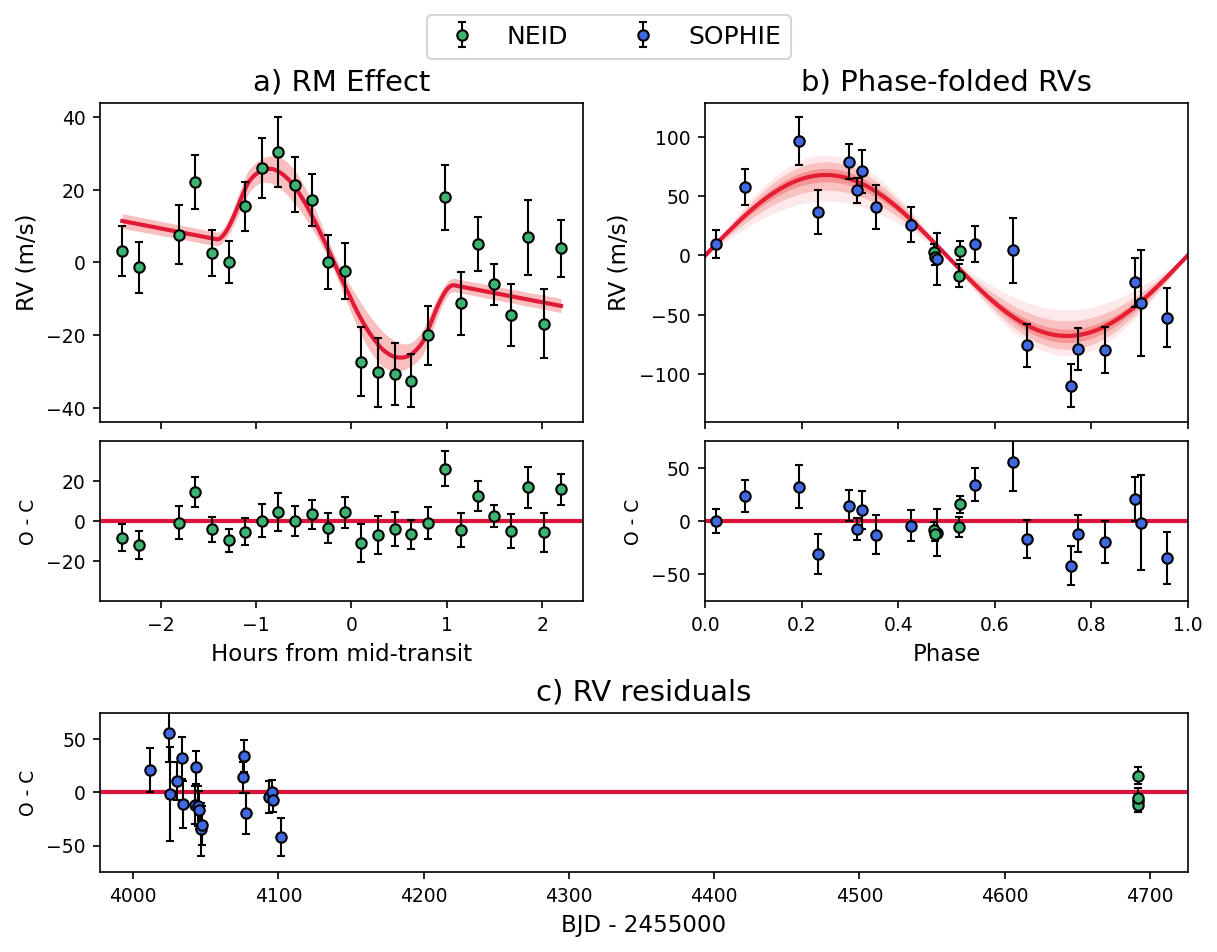}
\caption{Spectroscopic observations of TOI-1259A. a) NEID observations (in green) of the RM effect produced during the transit of TOI-1259Ab along with the best fit model in red. Residuals are shown below. b) The phase folded observations from SOPHIE (in blue), as well as out-of-transit NEID data, of TOI-1259Ab, with best fit model in red and the residuals below. c) Residuals of all the RV data and best fit model, over the time they were observed.}
\label{fig:RM_fits}
\end{figure*}

\begin{figure}[h!]
\includegraphics[width=0.95\linewidth]{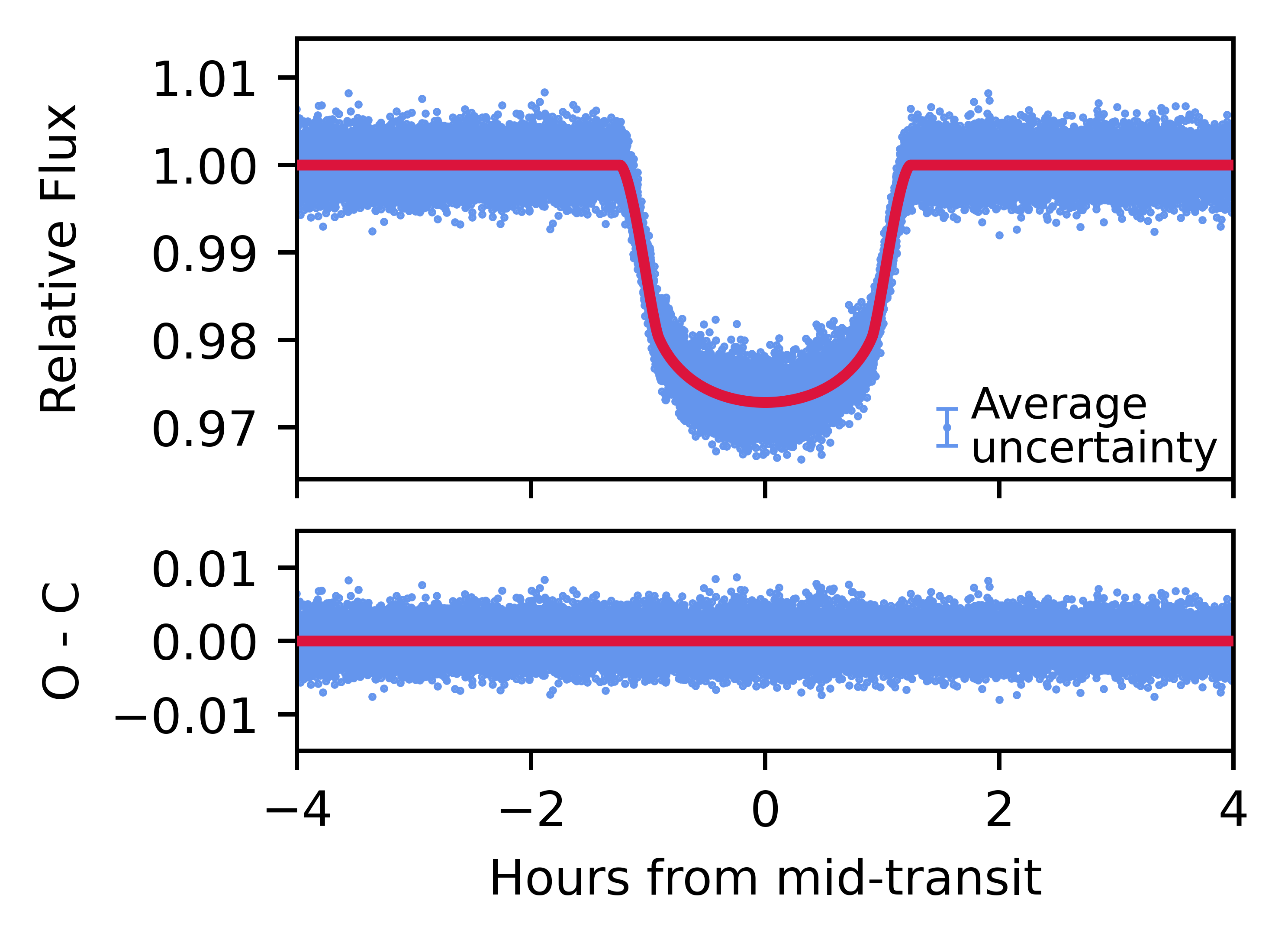}
\caption{Phase folded, detrended TESS data (blue) from the 30 TESS sectors analyzed in this work, along with the best-fit transit model from the \texttt{ironman} joint-fit analysis (red line). Data points are shown without error bars, but we show the median error of 2100 ppm. The residuals are shown in the bottom panel.}
\label{fig:TESS_fit}
\end{figure}

To derive the parameters of the planet and its orbit, we jointly modeled all the observations described in Section \ref{sec:observations} using the publicly accessible \texttt{ironman}\footnote{\url{https://github.com/jiespinozar/ironman}} package \citep{juan_polar_neptunes}. To model transit light curves, \texttt{ironman} employs \texttt{batman} \citep{batman_used_ironman}. To model RVs, \texttt{ironman} uses \texttt{rmfit} \citep{stefansson2020_rmfit_method,stefansson2022_rmfit}, which utilizes \texttt{radvel} \citep{fulton_radvel} to model the Keplerian orbit and the equations derived by \cite{hirano_2010_framework_in_rmfit} to model the RM effect. \texttt{ironman} jointly fits the data using a Bayesian approach, employing Dynamic Nested Sampling with the \texttt{dynesty} package \citep{dynesty_used_in_ironman} to sample the posteriors. To derive the stellar inclination angle, we followed \cite{stefansson2022_rmfit} and used the $\lambda$, $\cos i_\star$, $P_{\mathrm{rot}}$, and $R_\star$ parametrization of the RM effect that is available within \texttt{ironman}. This performs the correct accounting for the correlations between $v\sin i_\star$ and the equatorial velocity of the star \citep[see][]{masuda_psi_inference}. By sampling the posteriors for the stellar inclination ($\cos{i_\star}$), radius ($R_\star$), and rotation period ($P_{rot}$), the projected rotational velocity ($v\sin{i_\star}$) is calculated to model the RM effect, and $\psi$ is derived at the end of the sampling as:
\begin{equation}
\label{eq:psi}
\cos{\psi}=\cos{i_{\star}}\cos{i} + \sin{i_{\star}}\sin{i} \cos{\lambda}.
\end{equation}

We placed informative Gaussian priors on the stellar mass, radius, rotational period and the planet's orbital period based on the results of \cite{discovery_Martin}. For almost all the rest of the parameters, we placed uniform priors, except for the orbital period and time of mid-transit that were precisely constrained in Section \ref{sec:TESS_text}. In addition, for the `intrinsic stellar line width' parameter $\beta$ of the RM model, which represents both instrumental broadening and macroturbulence, we followed \cite{stefansson2022_rmfit}, and assumed an instrumental broadening term of 2.5 km s$^{-1}$ based on the resolution of NEID, and adopted a macrotubulent velocity of 2.5 km s$^{-1}$ based on the relationships from \citet{Valenti2005}, considering $T_{\rm eff} = 4775$ K. We added the instrumental and macroturbulence broadening in quadrature and set that as our prior for $\beta$, adopting an uncertainty of 2 km s$^{-1}$. The list of priors and resulting posteriors is shown in Table~\ref{tab:priors}.

\setlength{\tabcolsep}{10pt}
\renewcommand{\arraystretch}{1.3}
\begin{table*}[ht!]
    \centering
    \caption{Priors and resulting posteriors from the \texttt{ironman} joint fit of the available TESS photometry, the NEID RM effect RVs, and the out-of-transit RVs from SOPHIE. $\mathcal{U}(a,b)$ denotes a uniform distribution between a and b. $\mathcal{N}$($\mu$, $\sigma$) denotes a Gaussian distribution with mean $\mu$ and standard deviation $\sigma$. 
    $\mathcal{TN}$($\mu$, $\sigma$, $a$, $b$) denotes a Truncated Gaussian distribution with mean $\mu$, standard deviation $\sigma$ while constrained between the $a$ and $b$ limits. 
    $\mathcal{LU}(a,b)$ denotes a log-uniform prior between $a$ and $b$. The nested sampling jump parameters are the parameters shown with associated priors, and the remaining parameters are derived from those posteriors.}
    
    \begin{tabular}{lllr}\hline\hline
    Parameter & Description & Prior & Posterior \\
    \hline
    $\lambda$ & Sky-projected stellar obliquity (deg) & $\mathcal{U}(-180,180)$  &     $6^{+21}_{-22}$\\ 
    $\psi$ & 3-dimensional stellar obliquity (deg)  & ...  &      $24^{+14}_{-12}$\\ 
    $R_{\star}$ & Stellar radius ($\textrm{R}_\odot$)& $\mathcal{N}(0.711,0.020)$  & $0.715^{+0.015}_{-0.014}$\\     
    $M_{\star}$ & Stellar mass ($\textrm{M}_\odot$) & $\mathcal{N}(0.74,0.06)$  &   $0.748^{+0.048}_{-0.043}$\\
    $P_{\rm rot, \star}$ & Stellar rotational period (days)& $\mathcal{N}(28.0,2.8)$  &  $26.7\pm2.7$\\   
    $\cos i_{\star}$ & Cosine of stellar inclination & $\mathcal{U}(0,1)$  &  $0.25^{+0.21}_{-0.17}$\\    
    $v_{\rm eq, \star}$   & Stellar equatorial velocity $(\textrm{km\ s}^{-1})$ & ...  & $1.35^{+0.16}_{-0.13}$\\ 
    $v\sin i_{\star}$  &  Projected rotational velocity $(\textrm{km\ s}^{-1}$) & ...  & $1.28^{+0.16}_{-0.15}$\\ 
    $\rho_{\star}$   &  Stellar density $(\textrm{g\ cm}^{-3})$ & ...  & $2.89^{+0.03}_{-0.04}$\\ 
    \hline
    $R_{\rm{p}}$ & Planet Radius $(\textrm{R}_{\oplus})$ & ... & $11.6\pm0.2$ \\
    $a$  & Semi-major axis (\textrm{au}) & ... & $0.0408\pm0.0009$ \\
    $M_{\rm{p}}$ & Planet mass ($\textrm{M}_\oplus$) & ... & $132\pm12$ \\
    $\rho_{\rm{p}}$  & Planet density $(\textrm{g\ cm}^{-3})$& ... & $0.46\pm0.04$ \\ 
    $P$ & Orbital Period (days)& $\mathcal{N}(3.477978,0.000002$)  &      $3.4779792\pm0.0000001$\\  
    $t_0$ & Transit midpoint (BJD) & $\mathcal{N}(2458686.7005,0.0001$)  &     $2458686.70048\pm0.00004$\\ 
    $b$ & Impact parameter & $\mathcal{U}(0,1)$  & $0.11^{+0.04}_{-0.06}$\\   
    $R_{\rm{p}}/R_\star$ & Radius ratio  & $\mathcal{U}(0,1)$  & $0.149\pm0.004$\\ 
    $e$ & Orbital eccentricity & Fixed  &  0 \\
    %$\omega$ & Argument of periastron (deg) & Fixed(90.0)  &   90.0 \\
    $K$ & RV semiamplitude & $\mathcal{U}(0,1000)$ &   $68\pm6$\\    
    $a/R_\star$   &  Scaled semimajor axis & ...  &        $12.27^{+0.05}_{-0.06}$\\ 
    $i$   &  Orbital inclination    & ...  &         $89.5^{+0.3}_{-0.2}$\\ 
    \hline
    $q_1^{\rm TESS}$ & TESS linear limb-darkening parameter  & $\mathcal{U}(0,1)$  &   $0.41^{+0.03}_{-0.04}$\\ 
    $q_2^{\rm TESS}$ & TESS quadratic limb-darkening parameter & $\mathcal{U}(0,1)$  &   $0.36^{+0.03}_{-0.02}$\\ 
    $\sigma_{\rm TESS}$ & TESS photometric jitter (ppm) & $\mathcal{LU}(1,5\times10^{7})$  &  $4.4^{+9.2}_{-2.8}$\\
    $\gamma_{\rm SOPHIE}$ & SOPHIE RV offset $\textrm{(m\ s}^{-1})$ & $\mathcal{U}(-41000,-40500)$  &   $-40819\pm5$\\
    $\sigma_{\rm SOPHIE}$ & SOPHIE RV jitter $(\textrm{m\ s}^{-1})$ & $\mathcal{LU}$($10^{-3}$,100)  &  $0.5^{+12.4}_{-0.4}$\\  
    $q_1^{\rm NEID}$ & NEID linear limb-darkening parameter & $\mathcal{U}(0,1)$  &               $0.81^{+0.14}_{-0.26}$\\  
    $q_2^{\rm NEID}$ & NEID quadratic limb-darkening parameter & $\mathcal{U}(0,1)$ & $0.74^{+0.19}_{-0.37}$\\ 
    $\beta_{\rm NEID}$ & Intrinsic stellar line width $(\textrm{km\  s}^{-1})$ & $\mathcal{TN}(3.5,2.0, 0.0, 10.0)$  &  $0.94^{+2.15}_{-0.57}$\\  
    $\gamma_{\rm NEID}$ & NEID RV offset $(\textrm{m \ s}^{-1})$ & $\mathcal{U}(-50,50)$  &  $-2.2^{+1.8}_{-1.7}$\\ 
    $\sigma_{\rm NEID}$ & NEID RV jitter $(\textrm{m\ s}^{-1})$ & $\mathcal{LU}(10^{-3},100)$ & $0.7^{+5.2}_{-0.7}$\\  
\hline
    \end{tabular}
    \label{tab:priors}
\end{table*}

Figure \ref{fig:RM_fits} shows the results from the joint \texttt{ironman} fit for both the in-transit RM effect (Figure \ref{fig:RM_fits}a), and the out-of-transit RVs (Figure \ref{fig:RM_fits}b). The jointly-fitted TESS photometric data, along with the best model, are shown in Figure \ref{fig:TESS_fit}. The in-transit RVs show a clear RM effect signature of a prograde orbit, shifting towards positive RV (red-shifted) before reversing to negative (blue-shifted). From the fit, we find that TOI-1259Ab has a sky-projected obliquity of  $\lambda = 6^{+21}_{-22}\,\degree$, and a true 3D obliquity of $\psi = 24^{+14}_{-12}\,\degree$, which is compatible with an aligned orbit. The remaining best-fit parameter values listed in Table \ref{tab:priors} are in good agreement (within $1-2\sigma$ in all cases) with the parameters reported by \cite{discovery_Martin}.

Additionally, we experimented with fitting slightly different models to the available data. We evaluated the possibility of an eccentric orbit for the planet, and the possibility of having a long-term quadratic or linear trend (and all the possible combinations). We elected to adopt the circular fit without any long-term trend, as the Bayesian evidence ($\log{Z}$) favored this model over the other models we considered, having a Bayes factor $\Delta\log{Z}>2$ and $\Delta{\rm BIC}>10$ compared to all the other models.

% -----------------------------------------------------------------
% -----------------------------------------------------------------
% -----------------------------------------------------------------
\section{Discussion}
\label{sec:Discussion}
% -----------------------------------------------------------------
\subsection{TOI-1259Ab in Context}

\begin{figure*}
\centering
\includegraphics[width=\linewidth]{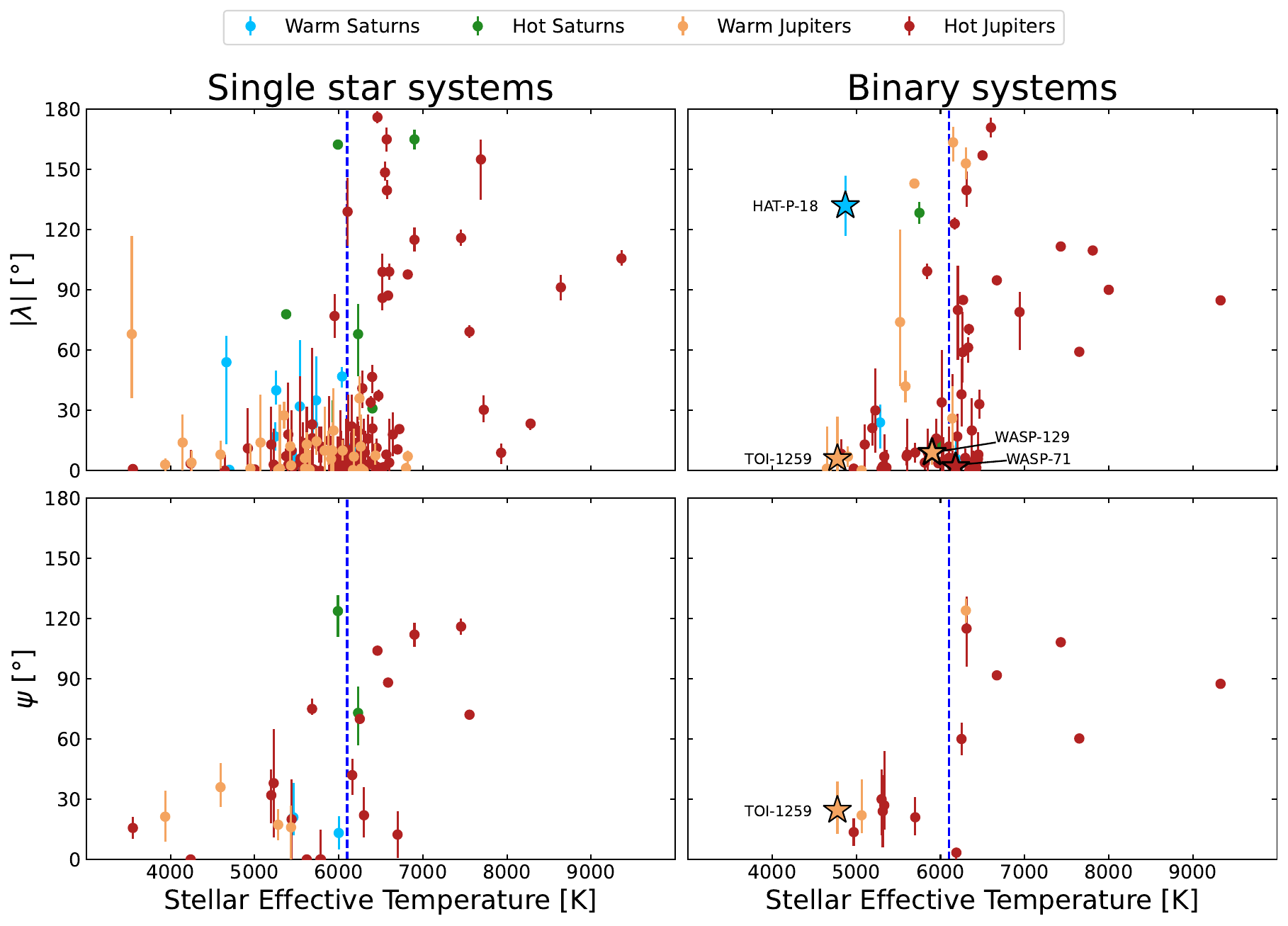}
\caption{Sky-projected obliquity $\lambda$ (upper panels) and 3-dimensional obliquity $\psi$ (bottom panels) as a function of the stellar effective temperature for the sample of planets around apparently single stars (left panels) and known binary/triple systems (right panels). Hot Jupiters (defined as having $0.4<M_{\rm{p}}/M_{\rm {J}}<13$ and $a/R_\star<11$) are shown in red, warm Jupiters ($0.4<M_{\rm{p}}/M_{\rm{J}}<13$ and $a/R_\star>11$) are shown in orange, hot Saturns ($0.2<M_{\rm{p}}/M_{\rm{J}}<0.4$ and $a/R_\star<11$) are shown in green, and warm Saturns ($0.2<M_{\rm{p}}/M_{\rm{J}}<0.4$ and $a/R_\star>11$) are shown in blue. Systems with known binary WD companions are shown as stars. The Kraft break at $\sim6100 \,\textrm{K}$ is shown as a dashed blue line. TOI-1259Ab is only one of the four WD systems with a measured true 3D obliquity.}
\label{fig:Population}
\end{figure*}

Figure \ref{fig:Population} shows the population of sky-projected (top) and 3-dimensional (bottom) obliquity measurements against the effective temperature of the planet's host star, separated into stars without known stellar companions (left) and known binaries or triples (right). The obliquities are taken from TEPCat\footnote{\url{https://www.astro.keele.ac.uk/jkt/tepcat/}} \citep{TEPCAT_paper} as of May 2025. At our time of retrieval of the TEPCAT database, a few published results
were missing and therefore added manually: KELT-25 and KELT-26 from \cite{kelt_25_26_martinez2020}, AB Pic b as reported by \cite{beta_pic_kraus2020spin}, TOI-451, TOI-455 and WASP-26 from \cite{19_obl_knudstrup2024} and WASP-35, WASP-44, WASP-45, WASP-54, WASP-91, WASP-99, WASP-129, WASP-162 and Qatar-7 from \citet{zak2025ten}. We excluded 9 systems (HAT-P-27, WASP-49, CoRoT-1, CoRoT-19, HATS-14, WASP-134, WASP-23, WASP-1, and WASP-2) 
that \citet{albrecht_dawson_winn_2022} proposed as being unreliable or too uncertain, but retained 55 Cnc\,e as a more recent measurement has been reported by \cite{55_cnc_zhao2023measured}.

We divided the sample into different types of planets using their NASA Exoplanet Archive\footnote{\url{https://exoplanetarchive.ipac.caltech.edu/index.html}} \citep{akeson2013nasa,christiansen2025nasa} entries. We considered four categories: hot Jupiters ($0.4<M_p/M_{\rm J}<13$ \& $a/R_\star<11$), warm Jupiters ($0.4<M_p/M_{\rm J}<13$ \& $a/R_\star>11$), hot Saturns ($0.2<M_p/M_{\rm J}<0.4$ \& $a/R_\star<11$), and warm Saturns ($0.2<M_p/M_{\rm J}<0.4$ \& $a/R_\star>11$). The NASA Exoplanet Archive also distinguishes apparently single stars from known multiple-star systems, but this is not fully up-to-date. We supplemented it by looking up the systems in four binary catalogs:
the common proper-motion catalogs based on Gaia DR3 \citep{Gaia_DR3} presented by \cite{million_binaries_Badry}, \cite{HAT_WD_finders}, and \cite{mugrauer2022gaia}, and the direct-imaging catalog of \cite{HAT_WD_reporters_measure}. 

Figure \ref{fig:Population} shows that the obliquity distributions of hot Jupiters around both single stars and binaries appear to be similar. They both exhibit a broadening of
possible obliquities at around $T_{\rm eff} = 6100$~K, moving from all being aligned, to more isotopically distributed as their hosts get hotter \citep{Winn2010, schlaufman2010evidence}. This turning point is consistent with the Kraft break \citep{Kraft1967}, below which stars have deep convective envelopes, perhaps allowing for more efficient tidal realignment than hotter stars,
which have radiative outer layers \citep[e.g.,][]{Winn2010,albrecht2012obliquities,Rice2022a,Zanazzi2024}. 

The picture for warm Jupiters is slightly different \citep{juan_8_warm_jupiters}. Around single stars, all the warm Jupiters in the sample are well-aligned \citep[e.g.,][]{Rice2022,Wang2024,juan_8_warm_jupiters}, while in binary systems the warm Jupiters show a wider range of obliquities. The alignment seen in warm Jupiters around single stars is probably a primordial feature, while the higher obliquities around binary stars are more consistent with high-eccentricity migration driven by the stellar companions \citep[e.g.,][]{Wu03,Fabrycky07}. With a WD companion at a projected separation of 1645 au, TOI-1259 (orange star in Figure \ref{fig:Population}) is well-aligned compared to a number of other misaligned warm Jupiters in binary systems. Further, TOI-1259Ab is only the third warm Jupiter in a binary system with a true 3D obliquity measurement. As discussed in Section \ref{sec:formation_pathways} below, despite having a prograde orbit, TOI-1259Ab is still consistent with formation via ELK migration induced by the WD companion.

A planet in a binary system where one of the stars is a WD is a relatively rare configuration among the sample of detected planets. \cite{WD_search_comp_exoplanet_host} found 2221 exoplanet-hosting binaries, of which only 17 have a WD companion, 10 of which were already compiled by \citet{discovery_Martin}. We find TOI-1259 to be one of four systems containing a known WD companion for which the stellar obliquity has been measured, with the other three systems being HAT-P-18 \citep{esposito_HAT_obliquity}, WASP-71 \citep{WASP-71b_Obliquity_smith2013, WASP-71_Obliquity_brown2017rossiter}, and WASP-129 \citep{zak2025ten}. However, the WD companions HAT-P-18B and WASP-71B were detected after the obliquity measurements were reported \citep{HAT_WD_finders, HAT_WD_reporters_measure,mugrauer2022gaia}. WASP-129, on the other hand, was found to have a WD companion by \citet{HAT_WD_finders}, after which \citet{zak2025ten} measured its stellar obliquity. Interestingly, TOI-1259Ab, HAT-P-18Ab, and WASP-129Ab are all warm gas giants, but show different obliquities. TOI-1259Ab, WASP-71Ab and WASP-129Ab are well-aligned with a sky-projected obliquities of $\lambda = 6^{+21}_{-22}\,\degree$, $\lambda = -1.9^{+7.1}_{-7.5}\,\degree$ and $\lambda = 9\pm6\degree$, respectively. In turn, HAT-P-18Ab is misaligned with a sky projected obliquity of $\lambda=132\pm15\degree$.

% ----------------------------------------------------
\subsection{Realignment and Circularization Timescales}
To investigate whether the observed properties of these WD systems are a result of primordial or post-formation processes, we began by estimating the timescales for tidal
realignment and orbital circularization.

Stars that are colder and positioned below the Kraft break have a tidal realignment timescale $\tau_{\rm r}$ that can be estimated as \citep{zahn1977_tidal_realign, albrecht2012obliquities}:
\begin{equation}
{\tau_{\rm r} =  10\times10^9 \left(\frac{M_{\rm p}}{M_\star}\right)^{-2} \left(\frac{a/R_\star}{40}\right)^{6}\,{\rm yr}}
\label{formula:Tidal_realignment}
\end{equation}

Using the results of the best fit discussed in Section \ref{sec:RM_fit_and_analysis}, we find a realignment timescale for TOI-1259Ab of $\sim3\times 10^4$ Gyr, which is significantly longer than the age of the universe \citep[e.g.,][]{lokhande2023estimating}, suggesting that tides have not significantly affected the obliquity.

To see whether TOI-1259Ab's orbit could have circularized after an eccentric formation process, similar to \cite{Juan_3362b_Obliq}, we follow \citet{Goldreich66} to estimate a tidal circularization timescale as:
\begin{equation}
{\tau_{c} \equiv -\frac{e}{\dot{e}} =  \frac{2P}{63\pi} Q'_p  \left(\frac{M_{\rm p}}{M_\star } \right) \left(\frac{a}{R_{\rm p}}\right)^5  F(e)}
\label{eq:circularization_initial}
\end{equation}
where $Q'_p$ is the planet's modified tidal quality factor \citep{Ogilvie07} and $F(e)$ is an eccentricity-dependent correction factor \citep{Hut81}. Note that $P$, $a$, and $e$ correspond to the initial values. Assuming that TOI-1259Ab started migrating with an initial eccentricity $e\,\sim\,0.9$ and pseudo-synchronous rotation of the planet, it yields $F(e) \sim 7 \times 10^{-6}$. Further, assuming $Q'_p = 10^5 - 10^6$, we estimate $\tau_{c} = 7 - 70$ Myr. This means the system could have easily circularized within the age of the system.

To put the tidal timescales in context of the other three WD-companion systems with an obliquity measurement, we also estimated the values for HAT-P-18, WASP-71 and WASP-129. For HAT-P-18, we find a tidal realignment timescale of $\sim10^6$ Gyr, which is so large it ensures the system has not started realigning, which is expected for a less massive planet on a relatively wide orbit that does not raise significant tides on its host star \citep[e.g.,][]{Louden2024}. In turn, its circularization timescale of $\sim10^2-10^3\,~\textrm{Myr}$ fits easily within the age of the system found by \cite{hartman_hat_18_age} to be $12.4\pm6.4$~Gyr, which means it could have undergone high-eccentricity migration driven by the stellar companion and still be observed in a circular and misaligned orbit today. 

For WASP-71, the picture is slightly more complicated. \cite{WASP_71_past_temp_triaud2011} showed that stars with masses $\geq 1.2\, \textrm{M}_\odot$, like WASP-71, start on the zero-age main sequence above the Kraft break and would presumably be less effective at realigning. But as \cite{WASP-71b_Obliquity_smith2013} noted, WASP-71 could have realigned after the star cooled and developed a convective envelope. However, even when assuming a convective envelope for the star's entire life, we find a realignment timescale of $8.1~\textrm{Gyr}$, which is longer than the age of the system of $\sim 3.5\, \textrm{Gyr}$ reported by \cite{WASP-71_Obliquity_brown2017rossiter}. Given its current (near) perfect alignment, we argue WASP-71 has not substantially realigned and formed well-aligned. Circularizing its orbit after eccentric migration would be possible on a much shorter timescale of $\sim 12-120\, \textrm{Myr}$.

Finally, for WASP-129 we estimate a tidal realignment timescale of $\sim3 \,\times \,10^4\, \textrm{Gyr}$ and a circularization timescale of $\sim \,1-10\, \textrm{Gyr}$, longer than the system's age of $\sim 1\,\textrm{Gyr}$ as reported by \citet{maxted2016five}. This makes it improbable that WASP-129Ab migrated from an eccentric orbit, suggesting a relatively quiescent history.

% -----------------------------------------------------------------
\subsection{$\gamma$ angle}
\begin{figure}
\centering
\includegraphics[width=0.9\linewidth]{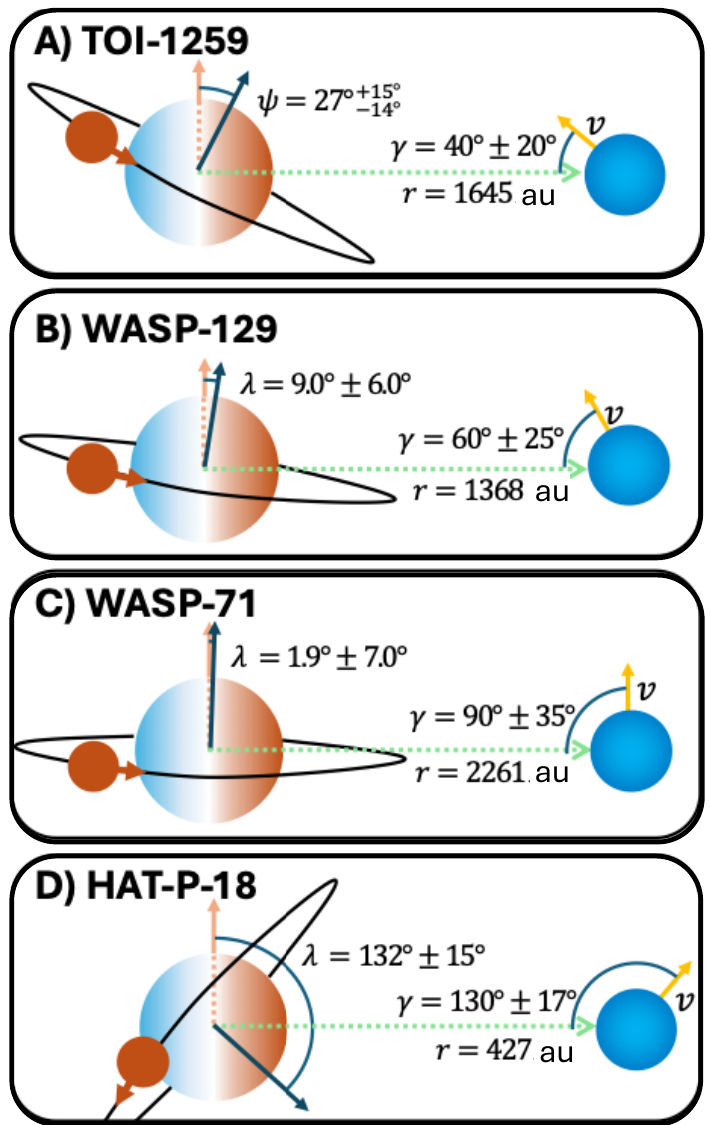}
\caption{A diagram depicting the relative angles of the WD binary systems. The obliquity $\psi$ (or $\lambda$ when only the sky projected obliquity is known) and $\gamma$-angle between the vector connecting the astrometric positions of the two stars ($\mathbf{r}$) and the difference in velocity vectors ($\mathbf{v}$) found in this study are shown, as well as the sky projected separations. The $\gamma$-angle is affected by the eccentricity and the relation to the orientation of the planet's orbit is unknown, meaning the $\gamma$-angle alone cannot constrain the orbits and is merely a proxy measurement. Distances and sizes are not to scale.}
\label{fig:system_angles_diagram}
\end{figure}

To shed further light on the possible formation mechanisms of TOI-1259Ab, we follow \cite{Rubenzahl2024} and constrain the angle $\gamma$ between the vector connecting the astrometric positions of the two stars ($\mathbf{r}$) and the difference in sky-projected velocity vectors ($\mathbf{v}$) using Gaia DR3 \citep{Gaia_DR3} positions and proper motions (Figure \ref{fig:system_angles_diagram}). As discussed by \cite{Rubenzahl2024}, any transiting planet has an orbital inclination $i\sim 90\degree$, and therefore, $\gamma\sim90\degree$ indicates a large misalignment between the planetary orbit and the binary orbit.
On the other hand, observing $\gamma\sim0\degree$ or $180\degree$ is indicative of good alignment, although the $\gamma$ angle is also affected by the eccentricity of the binary and the orientation of the line of nodes formed by the intersection of the planetary orbital plane and the sky plane. 

Following the formalism from \cite{Hwang2022}, for TOI-1259, we derive $\gamma=40\pm20 \degree$. Even with a high uncertainties in $\gamma$, the biggest barrier to identifying whether the binary orbit is aligned or misaligned with the planetary orbit are the unknown longitude of nodes in the system. As such, we argue---and further show in Section \ref{sec:formation_pathways}---that there is room for the ELK mechanism to operate as the WD companion is a potentially inclined perturber that could drive the cycles that can cause the planet to migrate.

In the other three WD systems, HAT-P-18, WASP-71 and WASP-129, the picture is clearer. The HAT-P-18 system has a $\gamma = 130\pm15\degree$ \citep{HAT_gamma}, we estimate $\gamma=90\pm 35\degree$ for WASP-71, and find $\gamma=60\pm 25\degree$ for WASP-129. In all cases, the binary and planetary orbits are highly misaligned, suggesting that the WD companion could be in a favorable configuration to drive ELK cycles. All system configurations are shown in in Figure \ref{fig:system_angles_diagram}.

% -----------------------------------------------------------------
\begin{figure*}[h!]
\centering
\includegraphics[width=\linewidth]{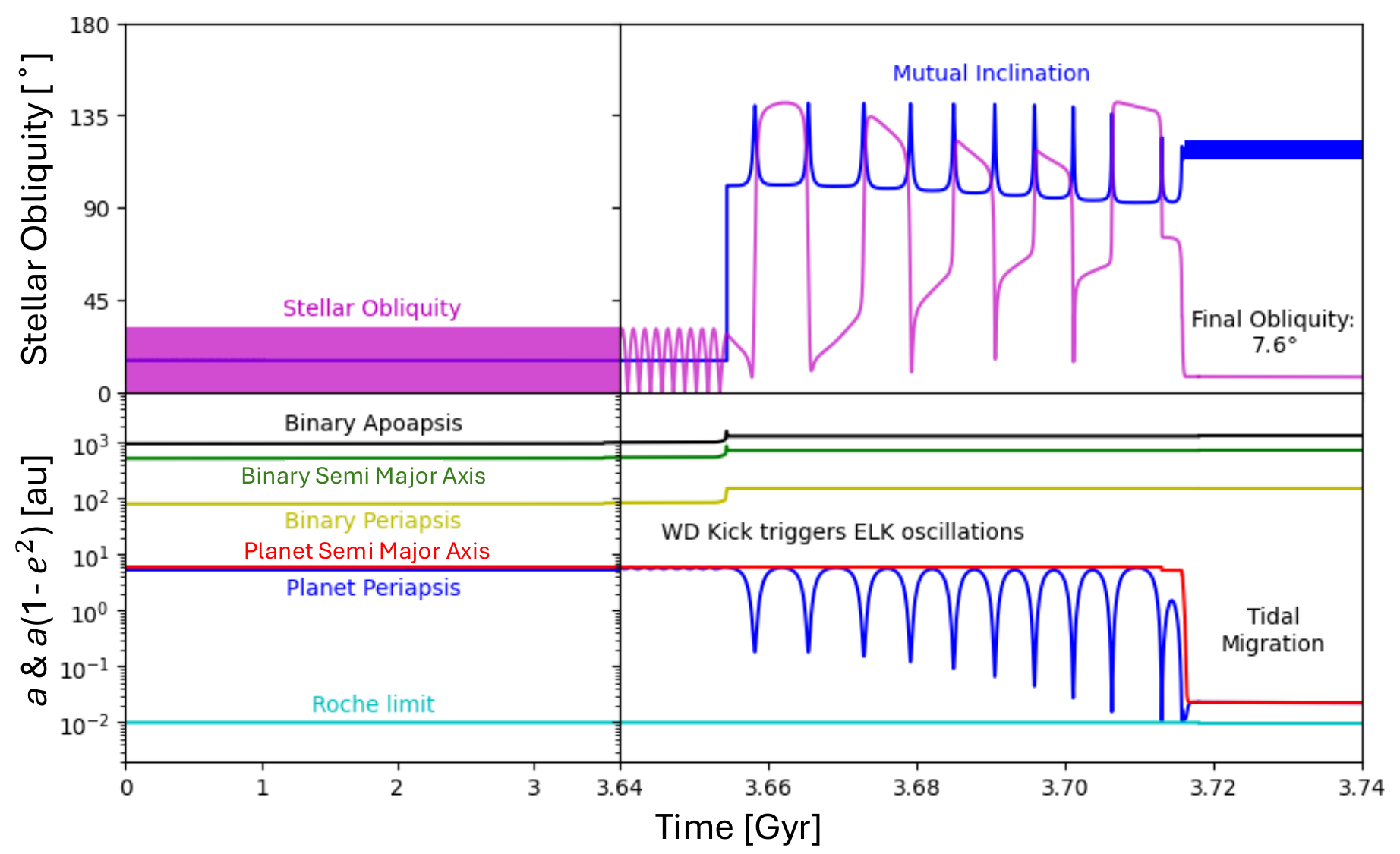}
\caption{Example evolution of a well-aligned gas giant in a binary system consistent with the TOI-1259 system. In this system, ELK oscillations are initially absent because of the low mutual inclination between the planetary and binary orbits (upper panels, blue line). The stellar obliquity (upper panels, magenta line) initially oscillates due to precession of the planetary orbit's line of nodes around the total system angular momentum. However, the WD kick changes the orbital parameters of the system and triggers strong ELK oscillations (lower right panel), causing high eccentricities for the planetary orbit and small periapsis distances (lower panels, blue line), eventually leading to shrinking of the planet's semimajor axis (lower panels, red line) via tidal migration. In this simulation, the planet's final obliquity is $7.6\degree$, and the final mutual inclination between the planetary and binary orbits is high (and oscillatory, as the planetary orbit
normal precesses around the star's spin axis due to tidal coupling). The WD companion has a wide final orbit with an apoapsis distance (lower panels, black line) of about $1300$~au, consistent with the projected separation of the TOI-1259 binary.}
\label{fig:sim_example}
\end{figure*} 

\begin{figure}
\centering
\includegraphics[width=0.9\linewidth]{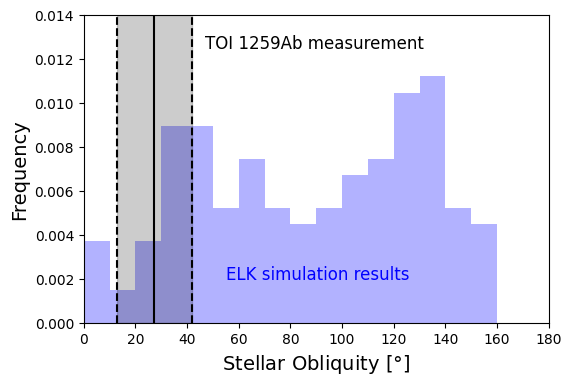}
\caption{Distribution of stellar obliquities for hot/warm Jupiters from our ELK simulations (blue histogram), compared to the measured obliquity of TOI 1259Ab including $1\sigma$ uncertainties (grey shaded region). About 14\% of the simulated close-in Jupiter systems have final obliquities consistent with the measured obliquity (19\% if any systems with obliquities lower than our measurement are included).}
\label{fig:obliquity_distribution_EKL}
\end{figure}

\subsection{Formation Pathways}\label{sec:formation_pathways}
With TOI-1259Ab being on a well-aligned or slightly misaligned orbit without the ability to tidally realign, there are many mechanisms that could explain its obliquity. Given that, it is difficult to decisively trace the system's obliquity to primordial or post-formation misalignment, or a combination of both. We briefly comment on these possibilities below.

\textit{Primordial misalignment:} The aligned orbit of TOI-1259Ab could be a relic of a primordial alignment between the star and the protoplanetary disk in which the planet formed and possibly underwent disk migration to arrive at the configuration we currently observe  \citep[e.g.,][]{Goldreich80,Lin86,Ward97}. The small degree of misalignment compatible with the data could have been produced by tilting the disk due to gravitational torques from the binary stellar companion TOI-1259B \citep[e.g.,][]{Batygin2012,Lai2014,Zanazzi2018}, or by a magnetic torque between the inner disk and the host star, which could have given the star a slight inclination with respect to the initial planet-forming disk \citep{evolution_disk_magnetic_warping}.

\textit{Post formation:} A system like TOI-1259 could also form through high-eccentricity migration, and more specifically through the WD kick formation scenario presented by \cite{Stephan+2024}. In this scenario, giant planets form in distant orbits, experience ELK oscillations, and eventually undergo high-eccentricity migration after the WD kick tilts the WD's orbit away from the planet's orbital plane. The systems we observe---except WASP-129---have circularization timescales shorter than their ages, making this pathway plausible. With the WD and planet on inclined orbits, this could lead to a broad range of values for the resulting
stellar obliquity. In the observed sample, HAT-P-18 is misaligned, while TOI-1259 and WASP-71 are consistent with good alignment. As discussed further below, the outcomes of the ELK oscillations can result in a broad array of obliquities, suggesting that all of the systems would be compatible with the WD kick formation scenario. Indeed, \cite{saidel2025atmospheric} also concluded that if TOI-1259Ab was formed far out, ELK migration induced by the WD companion would be a plausible migration pathway based on the timescale of ELK oscillations being shorter than that of the pericenter precession.

% -----------------------------------------------------

\subsubsection{Obliquity Distribution Produced by the ELK Scenario}

In order to gauge how consistent the WD kick formation scenario is with our observation of good alignment for TOI-1259Ab, we conducted dynamical simulations of $10,000$ systems using the code of \citet{Stephan+2024}, with the simulation parameters tailored to the TOI-1259 system (see Fig.~\ref{fig:sim_example} for an example simulation that produces a system consistent with TOI-1259). For the simulation, we set the primary star's mass to $0.744$~M$_\odot$, and the planet's mass and radius to $0.441$~M$_{J}$ and $1.022$~R$_{J}$, respectively. The stellar companion's initial mass was chosen randomly between $1.4$ and $1.6$~M$_\odot$, such that it would evolve into an approximately $0.57$~M$_\odot$ WD consistent with observations of TOI-1259B \citep{Fitzmaurice_2022_WD}, as determined by the stellar evolution code {\tt SSE} \citep{Hurley+2000}. The planet's orbit was set to be initially circular, with an initial semimajor axis (a) chosen randomly between $1$ and $10$~au, and aligned with the primary star's spin. The companion star's orbit initially has a random eccentricity between $0$ and $1$, with an initial semimajor axis chosen randomly between $10$ and 10,000~au, following a log-Gaussian distribution as described by \citet{DM1991}. The inclination between the two orbits is chosen randomly following an isotropic distribution (uniform in cosine), and the arguments of periapsis are chosen randomly between $0$ and $360\degree$. Only configurations that are long-term stable and hierarchical enough for EKL to be relevant are chosen for the simulations \citep[see, e.g.,][]{Naoz2016}. Finally, the WD formation kick follows the empirically determined kick parameters from \citet{el_Badry_WD_kick}. The kick velocity is drawn from a Maxwellian distribution, with the typical kick velocity (peak of the distribution) being about $0.75$~km s$^{-1}$. As we are agnostic about the exact kick mechanism during WD formation, the kick direction is chosen randomly.

Our main interest in these simulations is in the obliquity distribution of the resulting hot and warm Jupiters that are formed. While it is generally expected that ELK produces mostly misaligned planets, previous works have shown that a non-zero fraction of systems produce nearly aligned systems \citep[e.g.,][]{Naoz+2012}. As the formation pathway of \citet{Stephan+2024} does include a WD kick, we chose to conduct new simulations to specifically test how likely it is for ELK with kicks to produce systems consistent with our observations for TOI-1259. The resulting population, which can be seen in Figure \ref{fig:obliquity_distribution_EKL}, is a combination of pre- and post-kick migration with a roughly even split and similar $\psi$ distributions each. We find that, indeed, about 14\% of our simulations produce final obliquities consistent with the measured obliquity for TOI 1259Ab. If we also include any systems with obliquities lower than our measurement, this number rises to 19\%. These results indicate that the ELK mechanism with WD kicks is very well able to produce short-period gas giants with low obliquities. 

% -----------------------------------------------------------------
% -----------------------------------------------------------------
% -----------------------------------------------------------------
\section{Conclusions}
\label{sec:Conclusion}
Using in-transit RV observations with the NEID spectrograph, we have demonstrated that TOI-1259Ab---a warm Jupiter orbiting a K star with a WD companion---has an aligned orbit with a sky-projected obliquity of $\lambda= 6^{+21}_{-22}\,\degree$. Combining the sky-projected obliquity constraint from the RM analysis with the stellar rotation period and the stellar radius yields a true 3D obliquity of $\psi = 24^{+14}_{-12}\,\degree$, compatible with a prograde orbit. We simulated 10,000 WD kick-induced evolutions of systems like TOI-1259, which resulted in an obliquity consistent with our observations in 14~\% of systems, making it a plausible formation pathway for this system. However, for TOI-1259Ab, we cannot rule out more quiescent formation via e.g., disk-driven migration. Along with HAT-P-18, WASP-71, and WASP-129, we find TOI-1259 to be the fourth system with a WD companion to have its obliquity measured, starting to form a new population to study the influence that stellar evolution has on the dynamical evolution of planetary systems.

\section*{Acknowledgements}
We thank the Red Worlds Lab research group at the University of Amsterdam for their thoughtful discussions and suggestions that strengthened the manuscript. J.I.E.-R. gratefully acknowledges support from the John and A-Lan Reynolds Faculty Research Fund, from ANID BASAL project FB210003, and from ANID Doctorado Nacional grant 2021-21212378. 

We thank the NEID Queue Observers and WIYN Observing Associates for their skillful execution of our observations. Data presented were obtained by the NEID spectrograph built by Penn State University and operated at the WIYN Observatory by NOIRLab, under the NN-EXPLORE partnership of the National Aeronautics and Space Administration and the National Science Foundation. The NEID archive is operated by the NASA Exoplanet Science Institute at the California Institute of Technology. Based in part on observations at the Kitt Peak National Observatory (Prop. ID 2022A-970114), managed by the Association of Universities for Research in Astronomy (AURA) under a cooperative agreement with the National Science Foundation. The WIYN Observatory is a joint facility of the NSF's National Optical-Infrared Astronomy Research Laboratory, Indiana University, the University of Wisconsin-Madison, Pennsylvania State University, Purdue University, and Princeton University. The authors are honored to be permitted to conduct astronomical research on Iolkam Du'ag (Kitt Peak), a mountain with particular significance to the Tohono O'odham. Data presented herein were obtained from telescope time allocated to NN-EXPLORE through the scientific partnership of the National Aeronautics and Space Administration, the National Science Foundation, and the National Optical Astronomy Observatory. The research was carried out at the Jet Propulsion Laboratory, California Institute of Technology, under a contract with the National Aeronautics and Space Administration (80NM0018D0004). 

This work was partially supported by funding from the Center for Exoplanets and Habitable Worlds. The Center for Exoplanets and Habitable Worlds is supported by the Pennsylvania State University, the Eberly College of Science, and the Pennsylvania Space Grant Consortium. Computations for this research were performed on the Pennsylvania State University’s Institute for Computational \& Data Sciences (ICDS).

This work has made use of data from the European Space Agency (ESA) mission Gaia processed by the Gaia Data Processing and Analysis Consortium (DPAC). Funding for the DPAC has been provided by national institutions, in particular the institutions participating in the Gaia Multilateral Agreement.

This research made use of the NASA Exoplanet Archive, which is operated by the California Institute of Technology, under contract with the National Aeronautics and Space Administration under the Exoplanet Exploration Program. %This research made use of Astropy, a community-developed core Python package for Astronomy \citep{astropy2013}.

This work used computational and storage services associated with the Hoffman2 Shared Cluster provided by UCLA Office of Advanced Research Computing’s Research Technology Group, and the resources provided by the Vanderbilt Advanced Computing Center for Research and Education (ACCRE).


\section{References}
\bibliographystyle{aa}
\bibliography{Citations}

\begin{appendix}
\section{NEID Radial Velocities}\label{app:rvs}

\begin{table}[ht!]
    \centering
    \caption{NEID Radial Velocities}
    \begin{tabular}{ccc}\hline\hline
    BJD & Radial Velocity (m s$^{-1}$) & Radial Velocity Uncertainty (m s$^{-1}$) \\
    \hline
2459691.743419 & 0.8 & 6.9 \\
2459691.750682 & -3.6 & 6.9 \\
2459691.768026 & 5.3 & 8.1 \\
2459691.775289 & 19.8 & 7.4 \\
2459691.782554 & 0.3 & 6.2 \\
2459691.789818 & -2.2 & 5.8 \\
2459691.797083 & 13.1 & 6.6 \\
2459691.804348 & 23.8 & 8.2 \\
2459691.811613 & 28.1 & 9.5 \\
2459691.818877 & 19.1 & 7.5 \\
2459691.826142 & 14.8 & 7.2 \\
2459691.833407 & -2.1 & 7.4 \\
2459691.840671 & -4.6 & 7.7 \\
2459691.847936 & -30.6 & 9.4 \\
2459691.855201 & -32.5 & 9.5 \\
2459691.862466 & -33.1 & 8.6 \\
2459691.869730 & -34.8 & 7.3 \\
2459691.876995 & -22.3 & 8.0 \\
2459691.884260 & 15.6 & 8.8 \\
2459691.891524 & -13.6 & 8.6 \\
2459691.898789 & 2.8 & 7.4 \\
2459691.906054 & -8.3 & 5.6 \\
2459691.913319 & -16.8 & 8.5 \\
2459691.920583 & 4.6 & 10 \\
2459691.927848 & -19.1 & 9.6 \\
2459691.935113 & 1.6 & 7.9 \\ 
\hline
    \end{tabular}
    \label{tab:NEID_data}
\end{table}

\end{appendix}

\end{document}